\documentclass[twocolumn, amssymb, prl, epsfig, superscriptaddress]{revtex4}
\usepackage[utf8]{inputenc}
\usepackage{color}
\usepackage[normalem]{ulem}
 
\usepackage{amssymb}
\usepackage{amsmath}

\usepackage{graphicx}
\usepackage{epsfig}
\usepackage[bookmarks,colorlinks=true,urlcolor=blue]{hyperref}
\usepackage{float}

\begin{document}


\title{Evidence of Charge Density Wave transverse pinning by x-ray  micro-diffraction}
\author{E. Bellec}
\affiliation{Laboratoire de Physique des Solides, CNRS, Univ. Paris-Sud, Université Paris-Saclay, 91405 Orsay Cedex, France}
\author{I. Gonzalez-Vallejo}
\affiliation{Laboratoire de Physique des Solides, CNRS, Univ. Paris-Sud, Université Paris-Saclay, 91405 Orsay Cedex, France}
\affiliation{Laboratoire d'Optique Appliquée, ENSTA ParisTech, CNRS,Palaiseau, France}
\author{V.L.R.  Jacques}
\affiliation{Laboratoire de Physique des Solides, CNRS, Univ. Paris-Sud, Université Paris-Saclay, 91405 Orsay Cedex, France}
\author{A.A. Sinchenko}
\affiliation{M.V. Lomonosov Moscow State University, 119991 Moscow, Russia}
\author{A.P. Orlov}
\affiliation{Kotel'nikov Institute of Radioengineering and Electronics of RAS, 125009 Moscow, Russia}
\author{P. Monceau}
\affiliation{Institut Néel, CNRS and Université Grenoble-Alpes, BP166, 38042 Grenoble Cedex, France}
\author{S.J. Leake}
\affiliation{European Synchrotron Radiation Facility, 71 avenue des Martyrs, 38043 Grenoble Cedex 9, France}
\author{D. Le Bolloc'h}
\affiliation{Laboratoire de Physique des Solides, CNRS, Univ. Paris-Sud, Université Paris-Saclay, 91405 Orsay Cedex, France}


\date{\today}

\begin{abstract}

Incommensurate charge density waves (CDW) have the extraordinary ability to display non-Ohmic behavior when submitted to an external field. The mechanism leading to this non trivial dynamics is still not well understood, although recent experimental studies tend to prove that it is due to solitonic transport. Solitons could come from the relaxation of the strained CDW within an elastic-to-plastic transition. However, the nucleation process and the transport of these charged topological objects have never been observed at the local scale until now. In this letter, we use in-situ scanning x-ray micro-diffraction with micrometer resolution of a NbSe$_3$ sample designed to have sliding and non-sliding areas. Direct imaging of the charge density wave deformation is obtained using an analytical approach based on the phase gradient to disentangle the transverse from the longitudinal  components over a large surface of a hundred microns size. We show that the CDW dissociates itself from the host lattice in the sliding regime and displays a large transverse deformation, ten times larger than the longitudinal one and strongly dependent on the amplitude and the direction of the applied currents. This deformation continuously extends across the macroscopic sample dimensions, over a distance 10 000 times greater than the CDW wavelength despite the presence of strong defects while remaining strongly pinned by the lateral surfaces. This 2D quantitative study highlights the prominent role of shear effect that should play a significant role in the nucleation of solitons. 
\end{abstract}


\maketitle


Topological objects have recently received particular attention because of their exceptional stability.  In magnetic materials for example, skyrmions are topologically-protected field configurations with particle-like properties that are promising for applications in electronic and spintronic devices\cite{PhysRevB.88.184422,Zhang2017,Romming636,PhysRevLett.117.087202}.
Those topological objects carrying a magnetic singularity  are now experimentally observed, created, and manipulated in many systems, including multiferroic materials\cite{Seki198}, ferroelectric materials\cite{ncomms9542}, and semiconductors\cite{Kezsmarki2015}.

\begin{figure}[h]
\includegraphics[scale = 0.415]{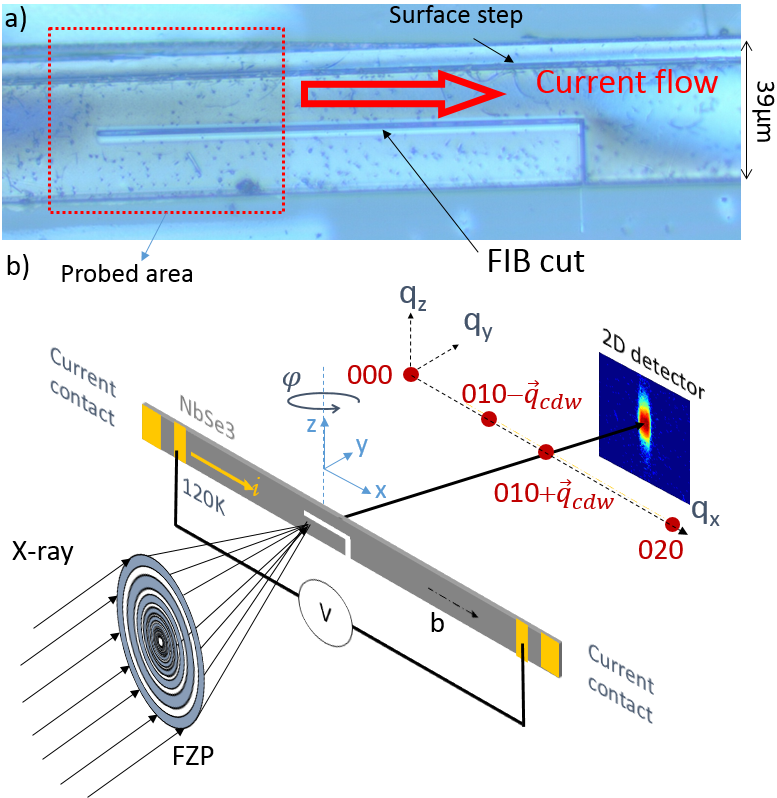}
\caption{a) Optical image of the NbSe$_3$ sample showing the line cut made by FIB, the presence of one step on the surface and the area probed by the focused x-ray beam with 1$\mu m$ steps. b) Schematic drawing of the experimental setup. The sample frame is rotated with respect to the lab frame by an angle $\varphi$. The CDW is along the $q_x$ direction, which corresponds to the $\vec{b}$ axis of NbSe$_3$.}
\label{setup}
\end{figure}

Topological objects carrying charges also play an important role in Charge Density Wave (CDW) systems. Indeed, when a sufficiently large current is applied to a sample displaying an incommensurate CDW, a collective charge transport is observed\cite{PhysRevB.85.241104,Monceau_review}. Those charges are carried by topological objects known as solitons, traveling through the sample over macroscopic distances, regularly spaced from each other\cite{PhysRevLett.52.663,PhysRevB.32.6582,PhysRevB.94.201120,PhysRevLett.100.096403, PhysRevB.85.035113} and periodically created in time \cite{PhysRevB.35.6348}.

This collective transport of charges is a complex phenomenon that requires a periodic creation and annihilation of solitons. 
This takes place close to the two electrodes, where the strain is maximum, and is based on elasticity and pinning of the CDW at the two edges of the sample: as a consequence, under an applied field,
the CDW is compressed at one edge and expanded at the other along the CDW wave vector.
This longitudinal CDW deformation has been observed by several indirect methods, like electromodulated
IR transmission\cite{PhysRevB.52.R11545} and local conductivity measurements\cite{Itkis_1993,PhysRevB.53.1833,PhysRevB.57.12781}.
The most direct measurement of this compression-dilatation phenomenon was obtained by x-ray diffraction,  also showing that beyond 100$\mu m$ from the two electrical contacts, the CDW wave vector varies little and linearly\cite{PhysRevLett.70.845,PhysRevLett.80.5631}. 
However, the transverse CDW deformations were not measurable in those two studies because the x-ray beam was larger than the sample width.

Even though the longitudinal deformation is directly associated to the nucleation of solitons, the transverse one also affects the collective current.
The first theoretical study dealing with shear deformations was reported in\cite{Elastic_Feinberg} where an elastic CDW submitted to an applied field and pinned by lateral surfaces display a quadratic phase behavior. From an experimental point of view, shear deformation influences transport properties: the threshold current increases with smaller sample widths and smaller cross sections \cite{small_o_TaS3,PhysRevB.46.4456,Size_dependent_threshold_fields}. The influence of shear on the collective current can be also observed through the response to current pulses \cite{PhysRevLett.114.016404}. From x-rays diffraction experiments, a transverse deformation around a surface step has been observed by topography in NbSe$_3$\cite{PhysRevLett.83.3514}, as well as a local transverse deformation of the CDW by coherent diffraction in TbTe$_3$ \cite{PhysRevB.93.165124} and in NbSe$_3$\cite{PhysRevLett.109.256402}. However, no qualitative and direct study dealing with both longitudinal and transverse deformations over a large area of the sample was reported despite its importance in the nucleation process.

In order to simultaneously measure the longitudinal {\it and} the transverse deformation, x-ray micro-diffraction experiment has been performed in an  NbSe$_3$ compound submitted to external $dc$ currents. Four gold contacts were evaporated on a 39$\mu m$ $\times$ 3$\mu m$ $\times$2.25$mm$ single crystal glued on a sapphire substrate to perform {\it in-situ} four point resistivity measurements. In addition, a L-shaped cut was made through the sample by a focused ion beam (FIB) so that the current could only flow in the upper half part of the sample, (see Fig. \ref{setup}a). This geometry allows to simultaneously observe the sliding area above the cut submitted to the currents and the non-sliding area from a single sample.

The connected sample was inserted into a cryostat mounted on the ID01 diffractometer at the ESRF and cooled to 120K, below the first CDW transition (Tc$_1 $=145K \cite {PhysRevB.18.5265,Hodeau_1978}). The satellite reflections associated with the CDW are located at $\pm\vec{q}_{cdw}$ of each Bragg reflection with an incommensurate wave vector $q_{cdw} = 0.243\pm 0.001 (\times \frac {2\pi} {b} $). The threshold current, measured {\it in situ}, was equal to I$_s $ = 0.5 mA  (see Fig. 2). The 8 keV X-ray beam was focused by a Fresnel zone plate down to 200nm$\times$300nm on the sample (see Fig \ref{setup}b)) and the diffracted intensity was recorded with a 2D Maxipix Detector with 55$\mu$m$\times$55$\mu$m pixel size, located 70 cm downstream. The Fresnel zone plate was mounted on a piezo stage to continuously map a 40$\mu$m$\times$90$\mu$m area, through the whole sample width, with a step size of 1 $\mu$m (see the probed area in Fig. \ref{setup} (a) and \cite{Chahine, StevenLeake} for more technical details). 
Rocking curves were performed at each position for the $Q_{020}$ Bragg and (0 $\vec {q}_{cdw}$ 0) satellite reflections for 6 currents, below and above the threshold and in the two opposite directions. We thus obtained a 5D intensity matrix containing the three reciprocal coordinates of the wave vector and the two spatial coordinates of the beam position.

\begin{figure}
\includegraphics[scale = 0.23]{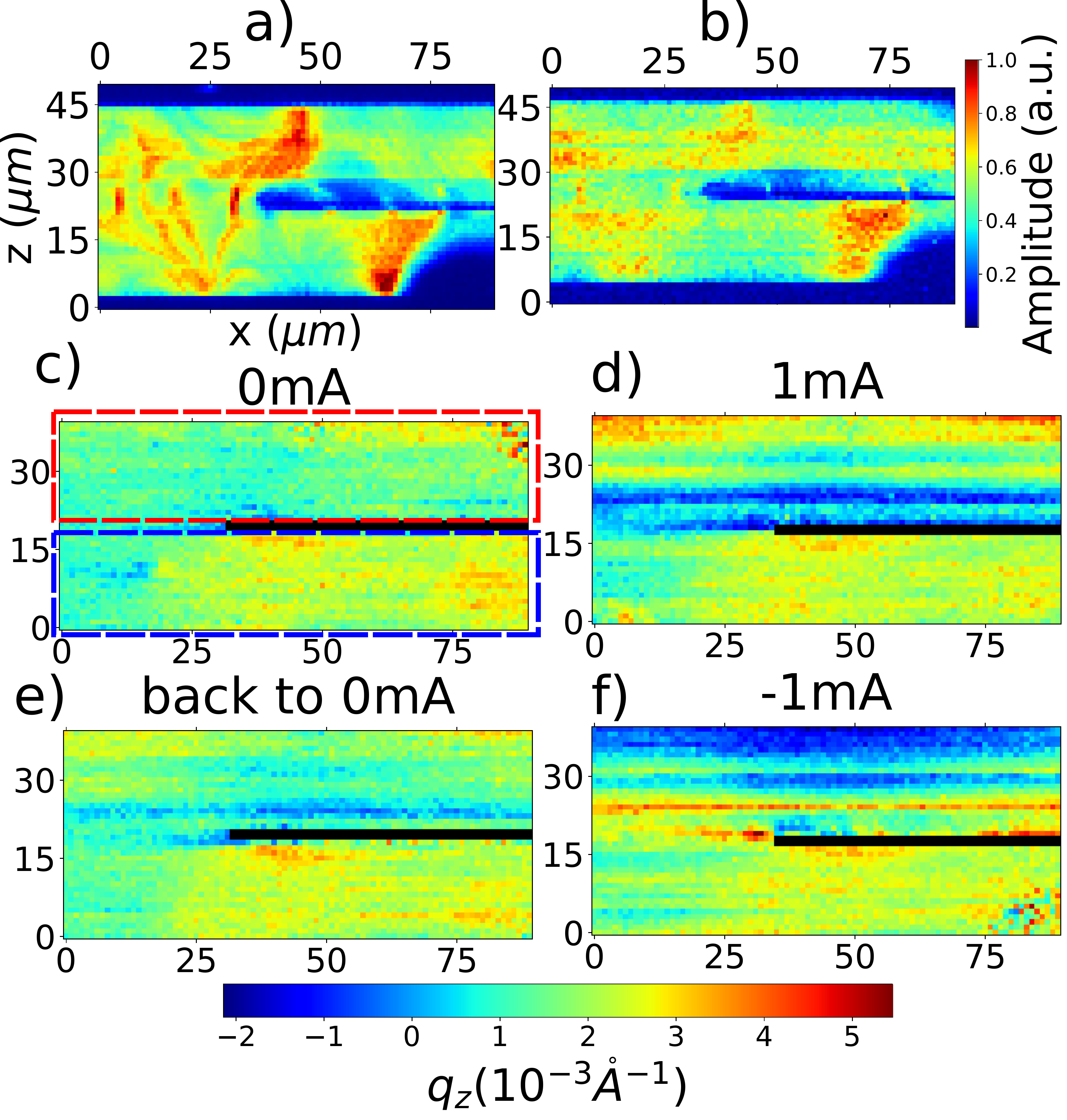}
\caption{a) Integrated rocking curve intensity for the (020) Bragg and b) of the $Q_s$=(010)+$\vec{q}_{cdw}$ as function of position around the pre-patterned line cut at I=-1mA. The absence of intensity at the line cut position is clearly visible, as well as a deformed region in the lower right corner of the map. Figures c) to f) display the $q_z$ transverse component of  $Q_s$ (the line cut has been removed for clarity). The maps c) to f) are scaled to  the sample width along z. The red and blue rectangles in c) correspond to the sliding and non-sliding area respectively.}
\label{kmap}
\end{figure}

 The $Q_{020}$ Bragg reflection  gives information on the host crystal lattice while the $Q_s=(0,1,0)+\vec{q}_{cdw}$  satellite reflection contains information from both the host crystal lattice and the CDW modulation. It is therefore crucial  to measure the two types of reflections to dissociate the CDW from the host lattice.
The correlation between the two is particularly visible on the maps displaying  the integrated intensity by summing over  the full rocking curve (see Fig.\ref{kmap} a) and b)).  The FIB cut is obviously visible on both maps, as well as disturbed areas of the host lattice reflecting on the CDW (see Fig. 4 in Supp. Mat.).  These perturbed areas remain weak, in the $\delta q=10^{-3}\AA^{-3}-10^{-4}\AA^{-3}$ range and are most probably due to constraints induced by the FIB at the line cut and at few specific places at the surface, leading to rocking curves extending beyond our measurement window  (see Fig. 2(a) and (b)).

The most interesting information comes from the wave vector orientation. The 3 coordinates of the two wave vectors are obtained independently from  Eq.1 in Supp. Mat.  from the  maximum of each fitted rocking curve ($\varphi$ angle) and the location of the maximum on the detector ($\delta$, $\gamma$).
Along the direction perpendicular to the sample surface (called the $y$ direction), the variations of the satellite $Q^y_{s}$ and the Bragg $Q^y_{020}$ are similar and small (see Fig.4 in Supp. Mat.).
Let us first consider the longitudinal $x$ deformation parallel to $\vec{q}_{cdw}$. As discussed in the introduction, a compression-dilatation of the CDW under current is expected but small far from electrical contacts. We indeed observed this small effect by averaging over the more homogeneous areas. Although $q_x$  is dominated by the host lattice variations, a small compression and dilation of the CDW period is well observed in the $10^{-4}\mathring{A}^{-1}$ range as expected, in agreement with \cite{PhysRevLett.70.845, PhysRevLett.80.5631}. Although weak, this variation is robust since the $q_x$ variation changes sign when the current is reversed in the sliding area while remaining constant in the non-sliding one (see Fig. 6 in Supp. Mat.). 

 The most surprising maps is the one displaying the coordinate $q_z$, the direction transverse to the CDW (see Fig. \ref{kmap} c-f).  
Unlike $q_y$ and $q_x$, the $q_z$ map is strongly different between the two reflections (see Fig. 4 in Supp. Mat.) and displays an unambiguous evolution with applied currents. At I=1mA, above the current threshold $I_{s}$, a clear distortion is visible in the sliding area. In this region, the amplitude of the shear deformation is 10 times larger ($\delta q_z \sim 10^{-3}\mathring{A}^{-1}$) than the longitudinal one ($\delta q_x \sim 10^{-4}\mathring{A}^{-1}$). Furthermore, the sign of $q_z$  changes with inverse currents: at I=-1mA, the deformation is opposite to the one at I=1mA (compare Fig. \ref{kmap} d) and f)), proving that the distortion is not due to sample heating. Going back to 0mA, the CDW does not relax to its ground state as in Fig. 2(c)). Hysteresis effects are a common feature observed by resistivity measurements in several CDW compounds\cite{PhysRevB.33.5858,MIHALY1984807} including NbSe$_3$\cite{PhysRevB.26.2298}. In contrast, the non-sliding area located below the line cut remains constant for all currents.

The CDW phase has been recovered from the diffraction patterns by using a phase gradient approach. This analytical method was preferred to Bragg
ptychography despite a weaker resolution\cite{cha:hal-01397101} because it is more appropriate to map large areas from weak satellite reflection by using a weakly coherent beam, about 10 times more intense than a fully coherent beam, without overlap between steps.
The CDW is described by a periodic lattice modulation $\rho(\vec{r})=A\cos(2k_fx +\phi(\vec r))$ where A is the CDW amplitude, $x$  is the direction parallel to the CDW and to the current direction (see Fig. \ref{setup}). $\phi(\vec r)$ is the phase that describes space-dependent CDW distortions. 
If we assume that the phase varies linearly between two steps, i.e. over 1$\mu m$,  the phase at $\vec{r} = (x,z)$ can be locally expanded to the first order $\phi(\vec{r}+\delta\vec{r}) \approx \phi(\vec{r})+\delta x\frac{\partial \phi}{\partial x}|_{\vec{r}}+\delta y\frac{\partial \phi}{\partial y}|_{\vec{r}}+\delta z\frac{\partial \phi}{\partial z}|_{\vec{r}}$. In that case, $q_{cdw}$ is nothing else but  the phase gradient\cite{PhysRevB.84.144109}. 
\begin{equation}\label{qtophi}
\delta q_i(\vec r) =\frac{\partial \phi}{\partial i}(\vec r)\ ,  \ i=x,y,z
\end{equation}
This first order approximation is justified if an elastic model of CDW deformation is considered where abrupt variations at the micrometer scale can not appear. It is also validated by the relevance of the results obtained afterwards. The estimation of the error made by this integration method is described in the Supp. Mat. It is however obvious that abrupt CDW phase shifts like CDW dislocations can not be observed.

\begin{figure}
\includegraphics[scale = 0.3]{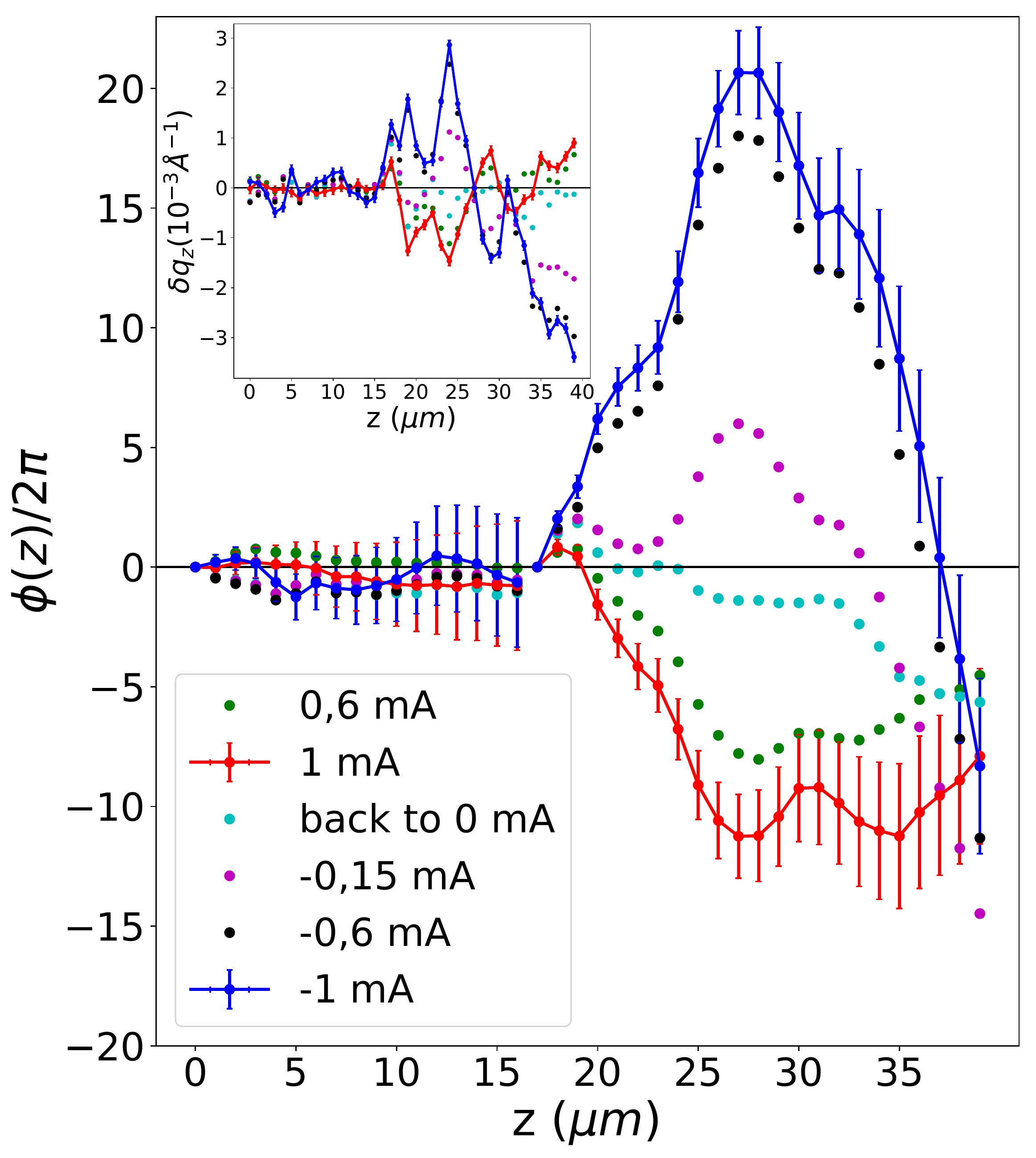}
\caption{Phase reconstruction of $\phi(z)$ averaged along the x direction for $\phi(z=17)=0$. The phase $\phi(z)$ is almost constant below the line cut, and displays a strong variation with currents in the other part. The sign of $\phi(z)$ is reversed when the current is reversed, keeping however the same value at the upper edge (z=39) of the sliding area. By this integration method, the error bars accumulate from left to right. They were only added on the $\pm 1$ mA curves for clarity. Inset : the corresponding $\delta q_z = q_z(I)-q_z(I = 0.15\text{mA})$ from which we reconstructed $\phi(z)$. The order of the legend corresponds to the chronological order of applied currents.}
\label{phase}
\end{figure}

\begin{figure*}
\includegraphics[scale = 0.35]{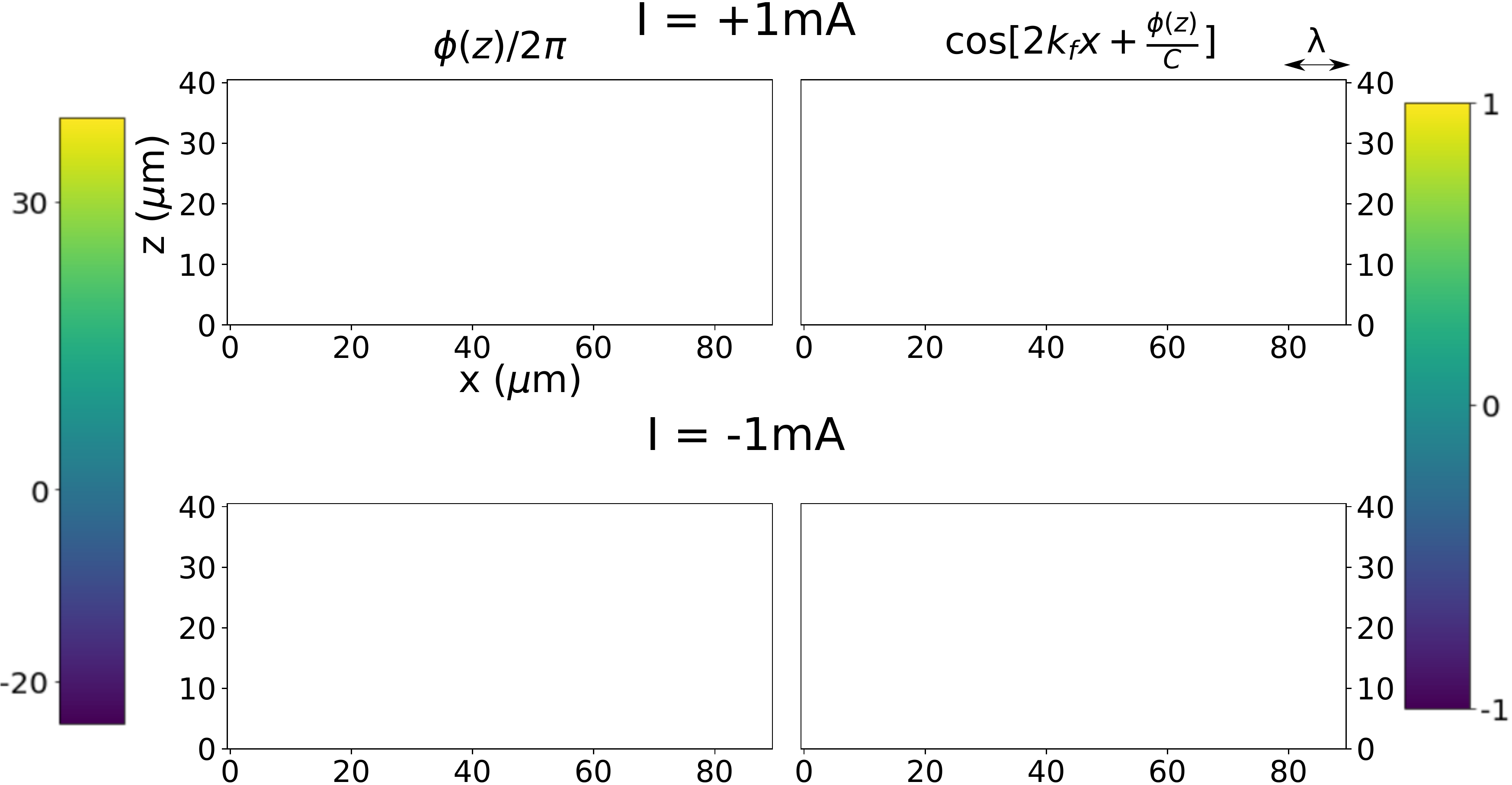}
\caption{2D reconstruction of the phase and of the CDW modulation around the  line cut for two inverse currents. For clarity, the CDW wavelength $\lambda$  is considerably increased to separate the wavefronts (in reality, $\lambda=14\mbox{\AA}$) and the phase $\phi$ is divided by an arbitrary constant C=210 in order to visualize the wavefronts.  Only the transverse component $\partial \phi/\partial z$ is considered, varying much more than $\partial \phi/\partial x$. Under the line cut, the CDW stays static as expected. Above the line cut, in order to minimize the free energy, the CDW tends to have a deformation along x, $\partial \phi/\partial x\neq 0$. But, pinning at the upper border and line cut fixes the phase there $\phi(z=\text{upper border})=\text{constant}$. To satisfy the two pinning centers, a distortion along z must take place, $\partial \phi/\partial z\neq 0$}
\label{schema}
\end{figure*}

The phase $\phi(\vec{r})$ is  obtained analytically by spatial integration of the measured $\delta q_z$. To avoid contribution of the crystal lattice distortion in the phase reconstruction (Fig. \ref{phase} and \ref{schema}), we subtracted all 5D matrices with the one obtained at I = 0.15$\text{mA}$  considered as the reference below the threshold. 

 Since $\delta q_z>>\delta q_x$, we can focus on the phase shear deformation $\phi(x,z)\approx \phi(z)$. The inset of Fig. \ref{phase} shows the variation profile $\delta q_z$ versus currents averaged over the $x$ direction. As expected from $q_z$ in Fig. 2., the CDW phase $\phi(z)$ remains constant for all currents in the non-sliding area ($z\in [0,17]$).  In contrast, the sliding area ($z\in [19,39]$) displays continuous and large variations.  Despite the presence of strong defects like the surface step (see Fig. 1(a)), the phase displays a remarkable continuity over 20$\mu m$, i.e. over 1.4 $10^4$ times the CDW wavelength ($\lambda=14\mbox{\AA}$). The $\phi(z)$ amplitude is maximum in the central part of the sliding region, reaching 20 CDW periods for I=-1mA. This CDW wavefront curvature seems impressive but it extends for 20$\mu m$, that is a phase shift $\Delta \phi(z)=2.8\ 10^{-3}(2\pi)$ per CDW period in average. The sign of the curvature depends of the current direction and the amplitude curvature increases with increasing currents. Furthermore,  $\phi(z)$ converges to a constant value for all currents at the upper edge (z=39$\mu$m) showing CDW pinning at the two lateral surfaces. Note that this constant value is obtained without any constraint in the integration procedure.
 
The map of $\phi(z)$ as well as the corresponding CDW modulation is displayed in Fig.4 for two opposite currents and for all currents in Fig.7 in Supp. Mat. It emerges from this measurement 3 unexpected points: 1. The large shear effect, ten times larger than the longitudinal one. 2. The pinning from the the two lateral surfaces and 3. the continuous deformation of $\phi(z)$ over the whole sliding area despite the presence of strong defects, especially the surface step visible on the sample image (Fig.1a). This collective transverse deformation is even more striking  at the the extreme left part of the map for x=30$\mu$m, where it starts to display a continuous deformation over the whole sample width, i.e. over 40 $\mu m$.

In order to explain  the appearance of the CDW shear, we introduce the phase part of the free energy containing the interaction with an applied electric field \cite{PhysRevLett.68.2066}:
\begin{align}\label{energy}
\begin{split}
F_{\phi} =&\int d^3\vec{r}[\frac{1}{2}A^2(K_x(\frac{\partial \phi}{\partial x})^2+K_y(\frac{\partial \phi}{\partial y})^2+K_z(\frac{\partial \phi}{\partial z})^2)\\
&-\frac{e\rho_s}{2k_f}U\frac{\partial \phi}{\partial x}]
\end{split}
\end{align}
where A is the CDW's amplitude, $K_i$ are the anisotropic elastic constants, $\rho_s$ is the condensate density and $U$ the applied potential ($E = -\vec{\nabla}U$).  Since the CDW in NbSe$_3$ is incommensurate, the free energy does not depend on $\phi$ but only on its derivatives. The last term is related to the fact that any phase deformation along x induces a local charge density $-\frac{e\rho_s}{2k_f}\frac{\partial \phi}{\partial x}$ \cite{gruner_book}. This last term differs in \cite{PhysRevLett.68.2066}.

When a positive electric field is applied (negative $U$ in Eq. (\ref{energy})), $\partial \phi / \partial x <0$ is favorable corresponding to a compression of the CDW wavelength as observed by diffraction \cite{PhysRevB.61.10640, 0295-5075-56-2-289}. However, shear effects ($\partial \phi/\partial z \neq 0$) are not energetically favorable in Eq. (\ref{energy})), unless one takes surface pinning into account. The only way to compress or expand a CDW ($\partial \phi / \partial x \neq0$)  while keeping the phase pinned on lateral surfaces is to add a shear deformation along z. The shear observed here is therefore nothing else but a consequence of the longitudinal CDW deformation and lateral surface pinning.

 The strong shear effect measured here explains why several resistivity measurements\cite{small_o_TaS3,PhysRevB.46.4456,Size_dependent_threshold_fields} showed that the threshold current strongly depends on the sample cross section. Finally, the nature of surface pinning remains unclear. A possible pinning mechanism may be due to small steps on the surface as proposed in \cite{Elastic_Feinberg} but a commensurabity of the CDW  at surfaces can not be excluded neither.

In conclusion, the understanding of solitons nucleation needs the  observation of the  CDW behavior under currents over a large enough area and a small enough resolution. For this, the CDW modulation, with ten angstroms period, has to be probed over hundred of micrometers with micrometer resolution. This has been achieved by coupling fast scanning x-ray microdiffraction and phase gradient method revealing that the CDW becomes independent of the host lattice, with a large CDW shear and a strong lateral pinning centers, while keeping its continuity over tens of micrometers. This dominant shear effect should plays an important role in the dynamics of solitons.


\begin{acknowledgments}
The authors acknowledge the ESRF for provision of synchrotron radiation facilities.
The work was partially supported by Russian State Fund for the Basic Research (No. 17-52-150007) and the French state (PRC CNRS/RFBR 2017-2019).
\end{acknowledgments}

\bibliographystyle{unsrt}
\bibliography{biblio_Ewen_Bellec}

\end{document}